\RequirePackage{lineno}
\documentclass{svproc}
\usepackage{url}
\usepackage{graphicx}

\newcommand{\pAu}   {p+Au}
\newcommand{\dAu}   {d+Au}

\newcommand{ \pT} {p_{\rm T}}

\newcommand{\sNN}{$\sqrt{s_{\rm NN}}$ }

\newcommand{\GeVc}{GeV/$c$ }

\begin{document}
\mainmatter              
\title{Recent results on Light Flavor from STAR}
\titlerunning{Recent results on Light Flavor from STAR}  
%
\author{Jie Zhao (for the STAR collaboration)}
\authorrunning{Jie Zhao (for the STAR collaboration)} 
%
\tocauthor{Jie Zhao (for the STAR collaboration)}
\institute{Purdue University, West Lafayette, IN, 47907, USA,\\
\email{zhao656@purdue.edu}
}

\maketitle              


\begin{abstract}
These proceedings present an overview of the recent results on light flavor by the STAR experiment at RHIC.

\end{abstract}
\section{Introduction}
Relativistic heavy-ion collisions are unique tools to study the properties of the quark-gluon plasma (QGP) in quantum chromodynamics (QCD)~\cite{RHIC}.
One important goal of the heavy-ion program in RHIC-STAR is to explore the QCD phase diagram~\cite{RHIC,BES}. 
At the RHIC top energy data were collected with different species from small to large collision systems,
allowing studies of the high temperature QCD medium to extract quantitative information on the QGP.
The Beam Energy Scan (BES) program with the collision energies from 7.7 to 64.2 GeV 
extended the studies to lower temperature and higher baryon densities on the QCD phase diagram.	
The main goal is to search for the turn-off of QGP signatures and signals of the first-order phase transition and the critical point~\cite{BES}. 
To further extend the coverage on the QCD phase diagram, the fixed-target program (FXT) is exploited to reach the higher baryon densities with the baryon chemical potential in the range of $\mu_{B} \approx$ 420-720 MeV.

Starting 2010 STAR has accumulated large volume data from 200 GeV down to 7.7 GeV. 
A rich body of results were produced pertinent to the properties of the QCD matter. 
In these proceedings, we highlight selected STAR results on light flavor measurements that
were presented in the "Strangeness in Quark Matter" 2019 conference. 
More details can be found in STAR related proceedings in Ref~\cite{SQM1,SQM2,SQM3}.

\section{Initial conditions}
\begin{figure}[htb]
  \begin{center}
    \includegraphics[width=0.6\textwidth]{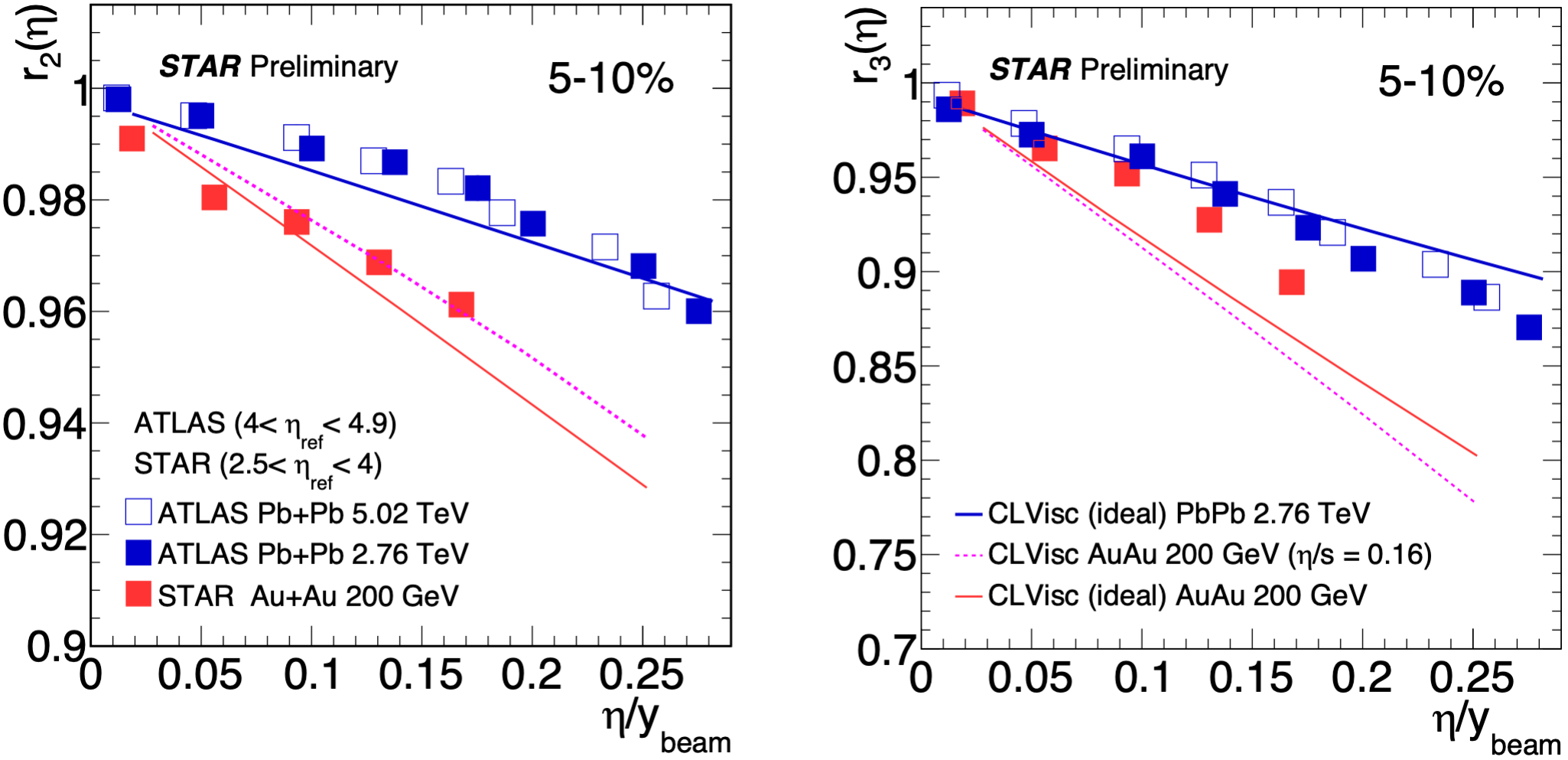}
	  \caption{Decorrelation parameters, $r_{2}$ (left), and, $r_{3}$ (right), as a function of the normalized rapidity in 5-10$\%$ Au+Au collisions~\cite{nie} and Pb+Pb collisions~\cite{dec1,dec2}.}
    \label{flowdec}
  \end{center}
\end{figure}

The measurement of longitudinal decorrelation of anisotropic flow can help provide a 3D image of the QGP evolution~\cite{dec0}.
Using the newly installed Forward Meason Spectrometer (FMS), 
STAR has measured longitudinal flow decorrelations in 200 GeV Au+Au collisions (Fig.~\ref{flowdec}). 
Such measurements as a function of the normalized rapidity, indicate a strong decrease with respect to  LHC~\cite{nie,dec1,dec2}. 
These results provide new constraints on both the initial-state geometry fluctuations and final-state dynamics of heavy-ion collisions.

\begin{figure}[htb]
  \begin{center}
    \includegraphics[width=0.8\textwidth]{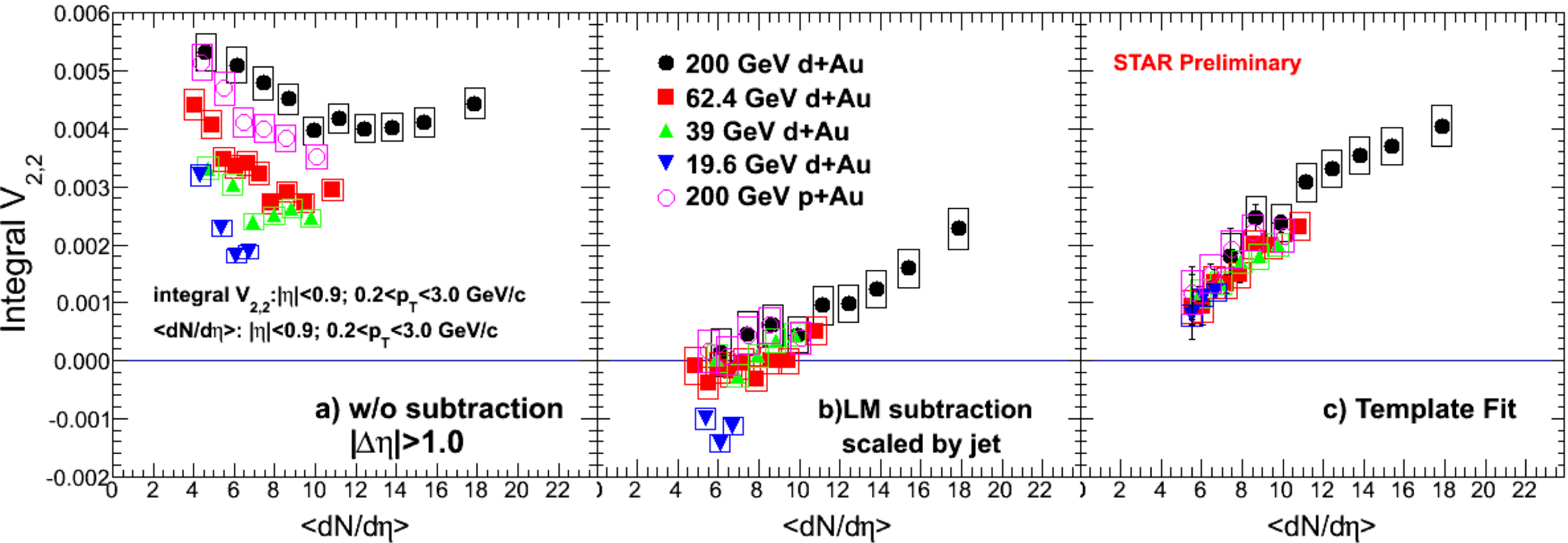}
	  \caption{Integral $\rm{V_{2,2}}$ as function of multiplicity in \pAu\ and \dAu\ collisions~\cite{huang}.}
    \label{flowsmall}
  \end{center}
\end{figure}

The measurement of the elliptic anisotropy ($v_{2}$) in small systems 
can further improve	our understanding of the importance of the initial geometry.  
	Figure~\ref{flowsmall} shows the $\rm{V_{2,2}}$ obtained by difference methods in \pAu\ and \dAu\ collisions~\cite{huang}.
The results for different energies show a common trend with the charged particle multiplicity, which provide important insights on the nature of collectivity in small systems.

\section{Phase transition and critical point}
\begin{figure}[htb]
  \begin{center}
    \includegraphics[width=0.32\textwidth]{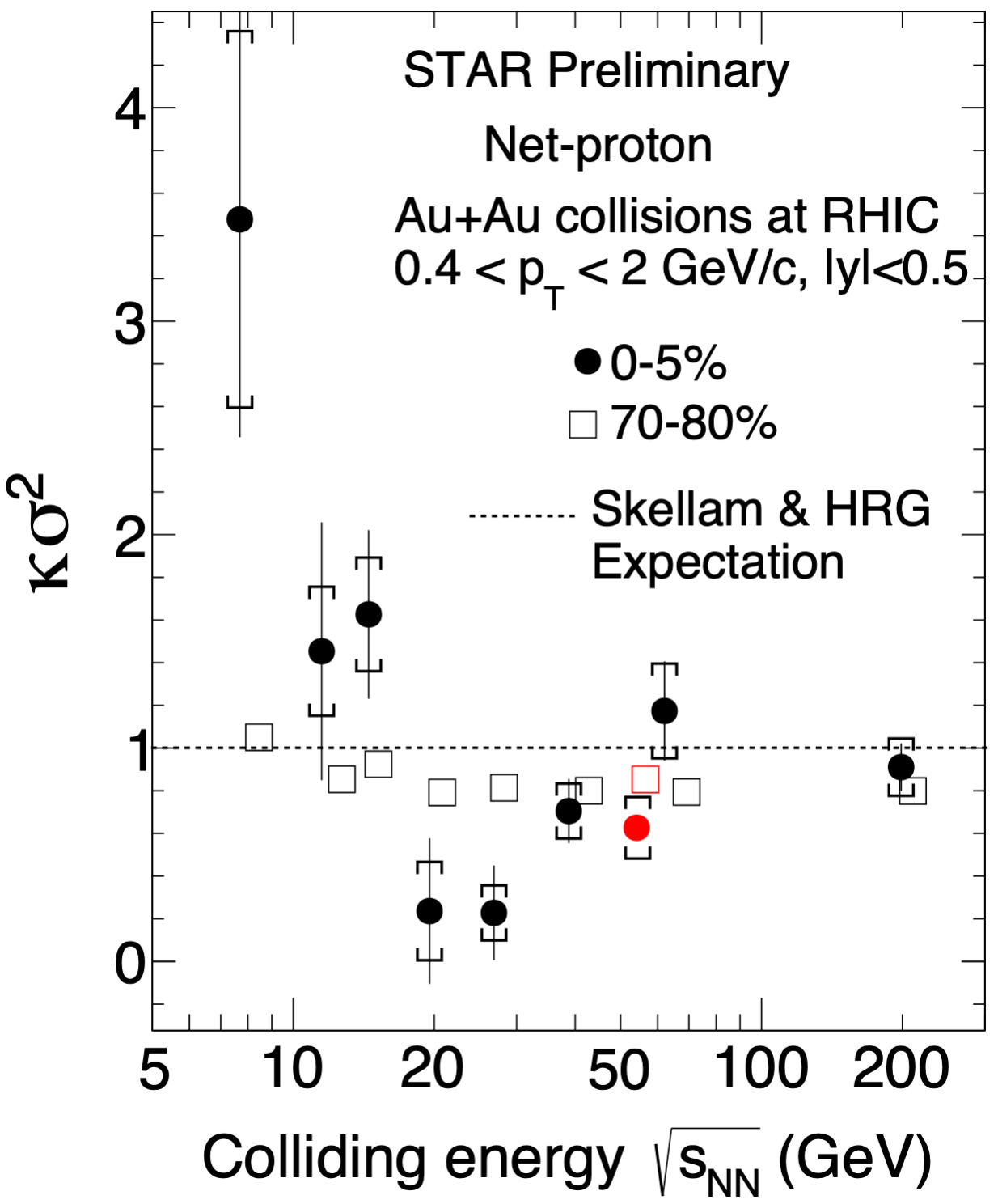}
    \includegraphics[width=0.59\textwidth]{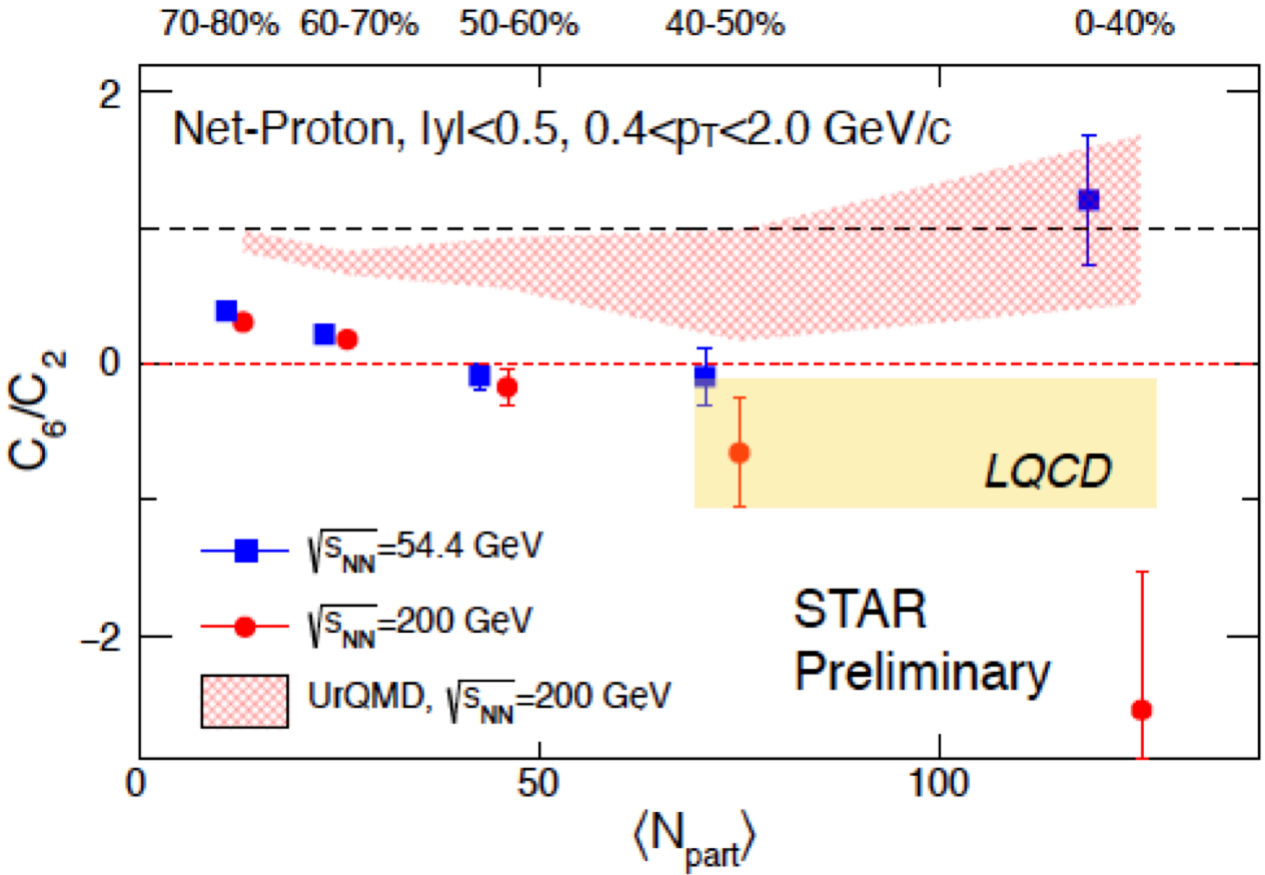}
	  \caption{Energy dependence of the net-proton moments products, $k\sigma^{2}$ (left), and the $6^{th}$- to $2^{nd}$-order cumulant ratio, $C_{6}/C_{2}$ (right), in 54.4 and 200 GeV Au+Au collisions~\cite{SQM1}.}
    \label{cumulants}
  \end{center}
\end{figure}

The higher-order fluctuation observables or the higher moments of conserved quantities, can be directly connected to the corresponding thermodynamic susceptibilities.
It is a sensitive tool to study the criticality on the QCD phase diagram as well as to determine the freeze-out parameters~\cite{netp1,netp2}. 
Figure~\ref{cumulants} (left) shows the new measurements of the net-proton cumulants in Au+Au collisions at 54.4 GeV~\cite{SQM1}. 
The data are compared to the results from other energies and a good agreement is found.
A non-monotonic behavior as a function of the collision energy is observed. 
Figure~\ref{cumulants} (right) shows the $6^{th}$- to $2^{nd}$-order cumulant of the net-proton multiplicity distributions~\cite{SQM1}. 
The $C_{6}/C_{2}$ for central Au+Au collisions at 54.4 GeV is positive while that for 200 GeV is negative,
although with large uncertainties. 
The results are in agreement with the theoretical expectation of a smooth crossover phase transition~\cite{c6A,c6B}.
STAR also measured net-$\rm{\Lambda}$ cumulants, which provide insights on the flavor dependence of the freeze-out parameters~\cite{SQM2,netL}.


\begin{figure}[htb]
  \begin{center}
    \includegraphics[width=0.45\textwidth]{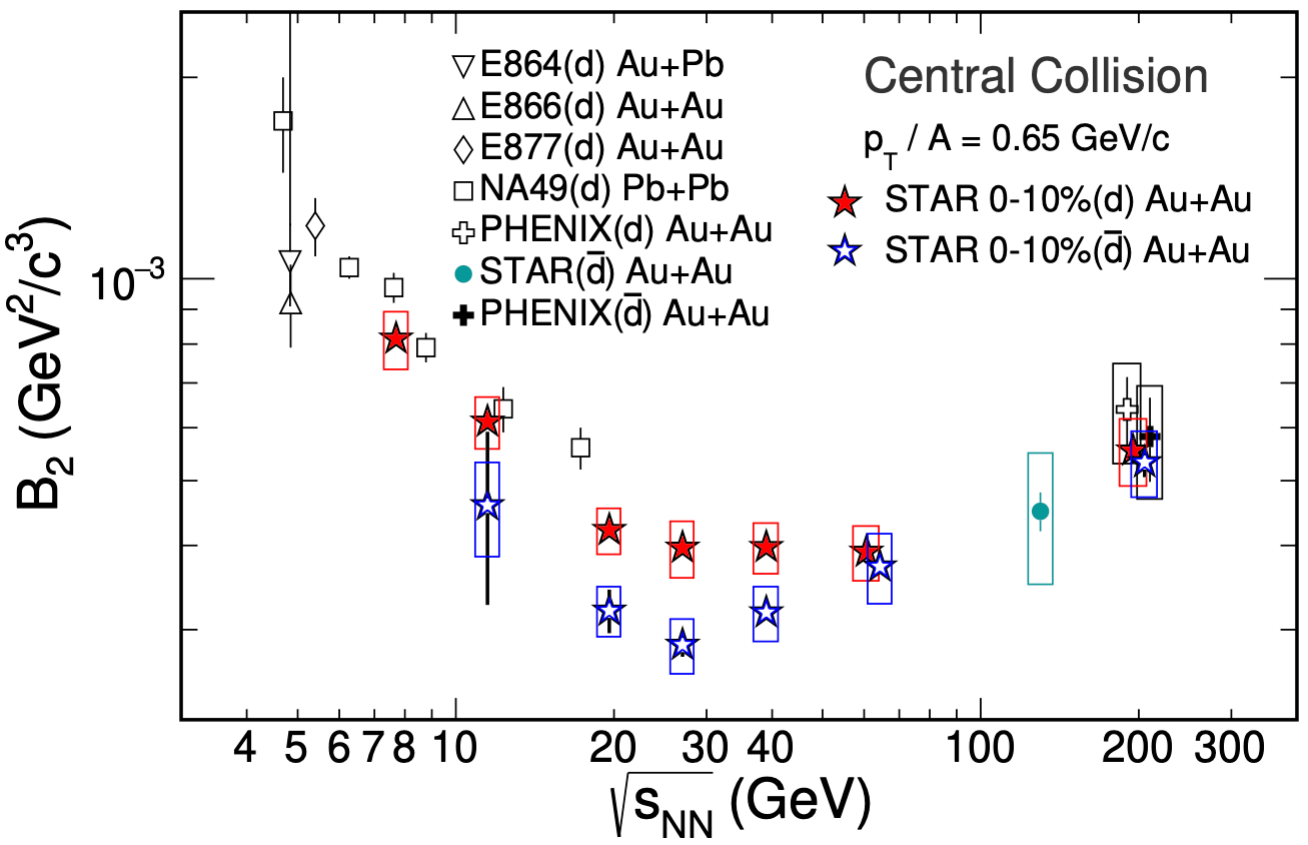}
    \includegraphics[width=0.40\textwidth]{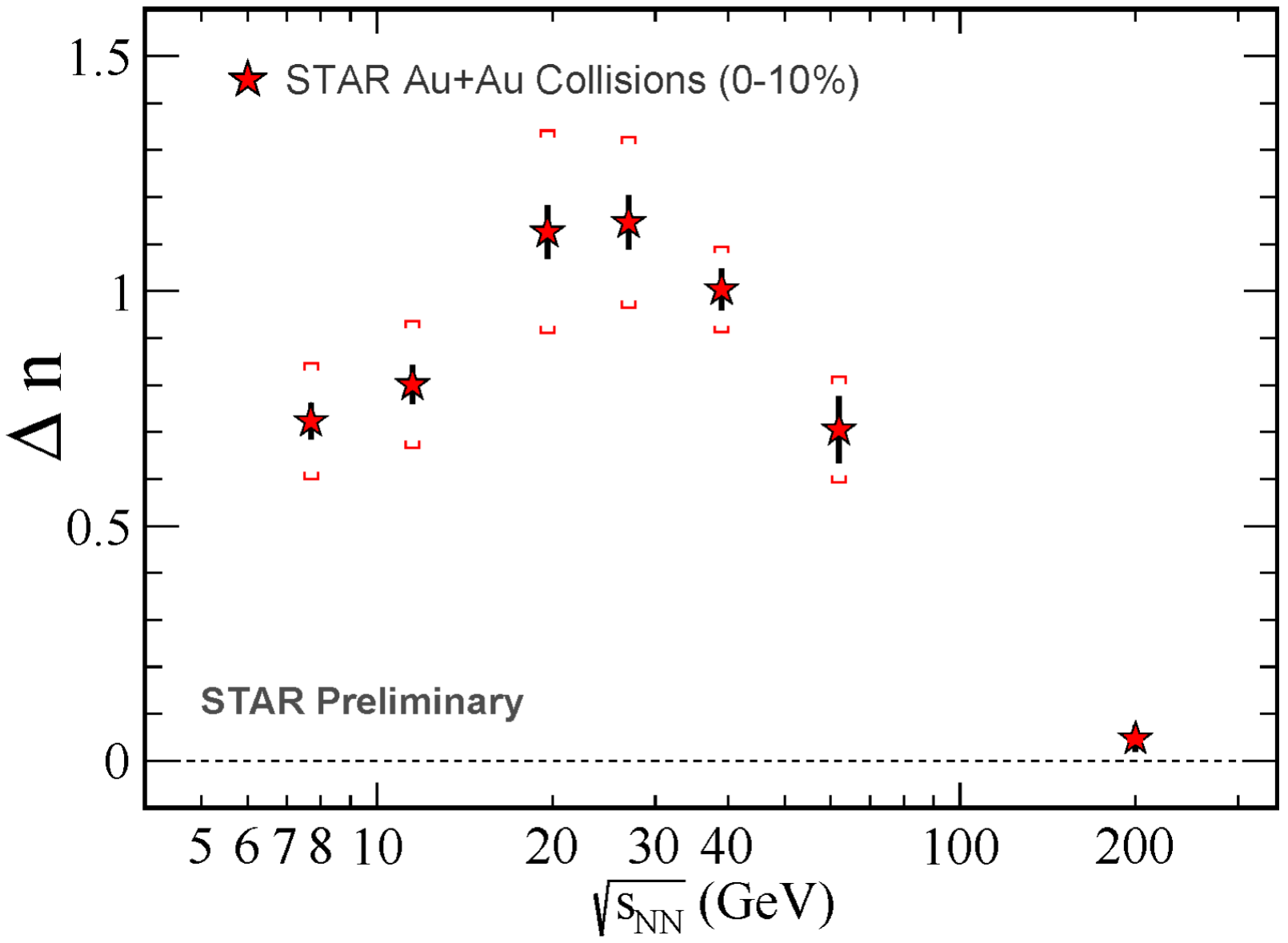}
	  \caption{Energy dependence of the coalescence parameter, $B_{2}$ (left), and the neutron density fluctuation, $\rm{\Delta}$n (right), from Au+Au collisions at RHIC~\cite{B2STAR,dNSTAR}.}
    \label{dTspectra}
  \end{center}
\end{figure}

Production of light nuclei with small binding energies, such as the triton ($\sim$8.48 MeV) and the deuteron ($\sim$2.2 MeV), 
formed via final-state coalescence, are sensitive to the local nucleon density~\cite{B2a}.
The production of these nuclei can therefore be used to extract information of nucleon distributions at freeze-out, 
which could be associated with the QCD phase transition~\cite{B2b}.
Figure~\ref{dTspectra} (left) shows that the coalescence parameter, $B_{2}$, first decreases and then increase with collision energy~\cite{B2STAR}.
The extracted neutron density fluctuations~\cite{dN}, $\rm{\Delta}$n, also show a non-monotonic behavior with collision energy (right panel of Fig.~\ref{dTspectra})~\cite{dNSTAR}. 

\begin{figure}[htb]
  \begin{center}
    \includegraphics[width=0.5\textwidth]{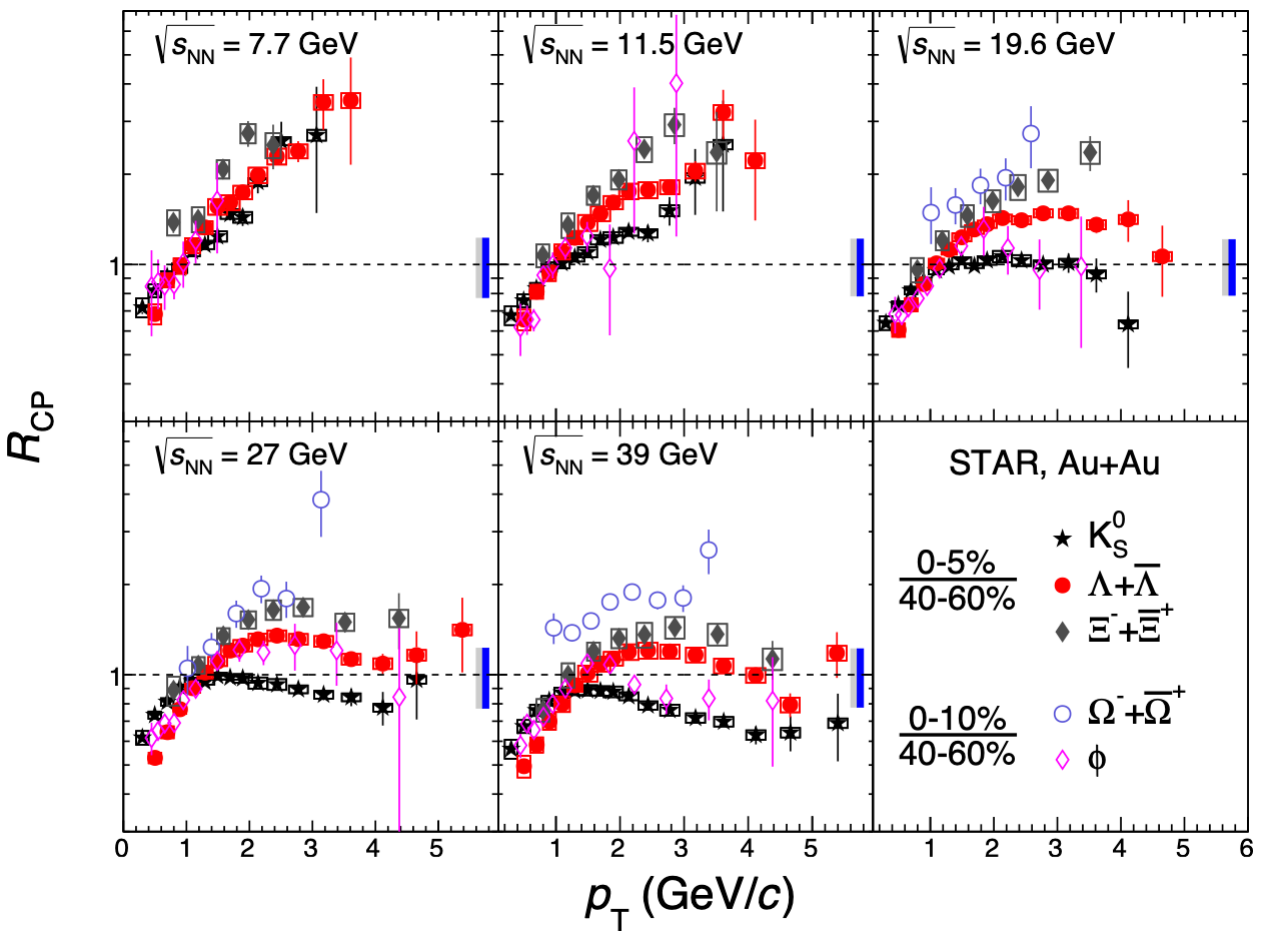}
	  \caption{$R_{cp}$ of the $K^{0}_{s}, \Lambda, \Xi, \phi, \Omega$ in Au+Au collisions at \sNN = 7.7 - 39 GeV~\cite{BESst}.}
	  \label{Rcp}
  \end{center}
\end{figure}

One of the important QGP signatures is the nuclear modification factor, $R_{CP}$, being significantly smaller than unity at high energies.
The strange hadron measurements from BES-$\rm{\uppercase\expandafter{\romannumeral1}}$ by STAR~\cite{BESst} show no suppression of the $K^{0}_{s}$ $R_{cp}$ up to $\pT$= 3.5 \GeVc. 
The particle type dependence of $R_{CP}$ is found to be smaller at \sNN $\leq$ 11.5 GeV (Fig.~\ref{Rcp}).
These measurements point to the beam energy region below 19.6 GeV for further investigation of the deconfinement phase transition.

\section{Hypertriton}
The measurement of hypertriton can provide insight on hyperon-nucleon interactions~\cite{Chen2018,Hyper2010}. 
The heavy-flavor tracker (HFT) significantly improved the signal-to-background ratio of hypertriton, 
thus allowing more precise determinations of the hypertriton binding energy and mass difference between hypertriton and antihypertriton. 
The STAR data~\cite{Hyper} provide the first test of the CPT symmetry in the light hypernuclei sector.
No deviation from the exact symmetry is observed.

\section{Medium effects and dynamics}
\begin{figure}[htb]
  \begin{center}
    \includegraphics[width=0.5\textwidth]{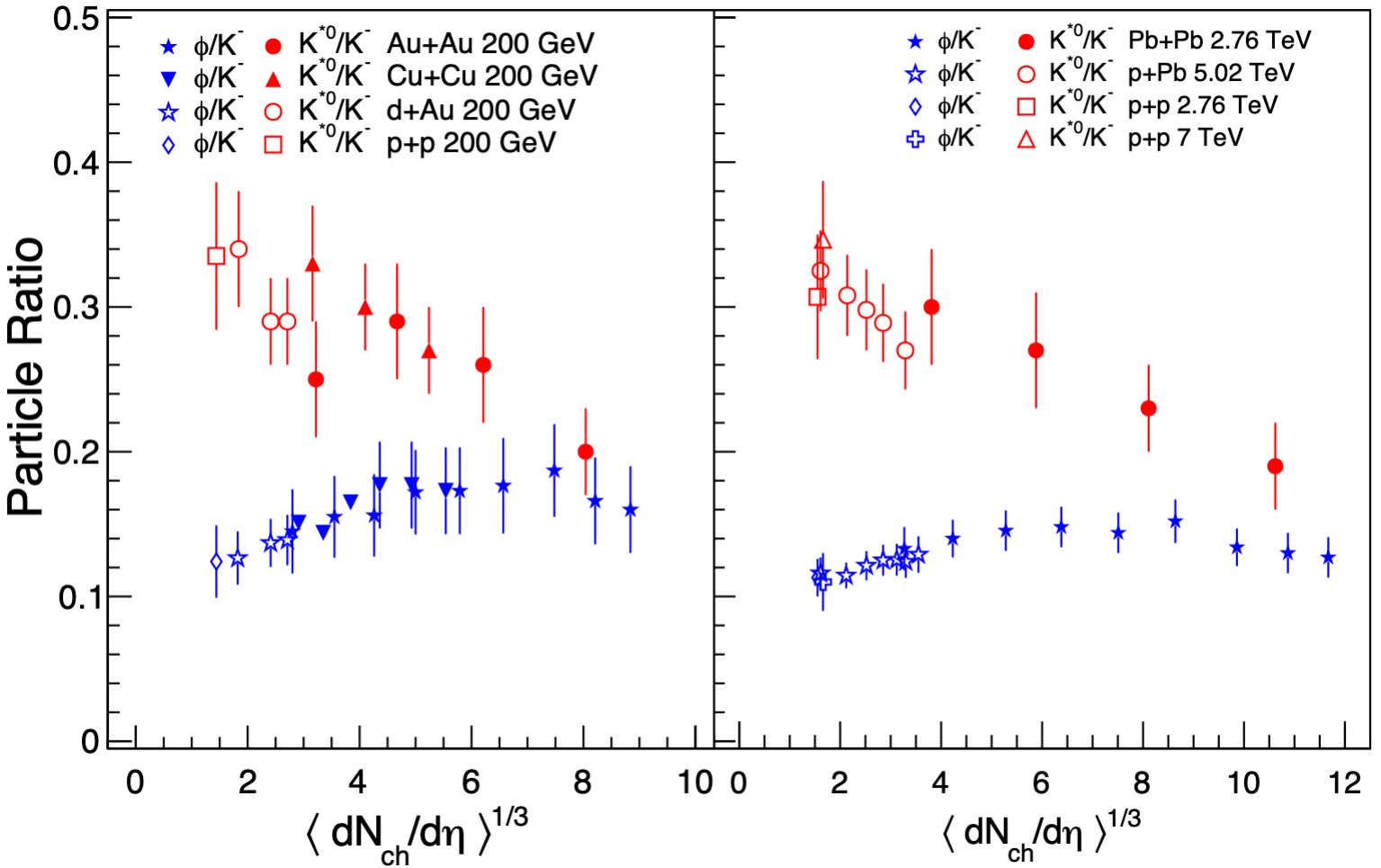}
	  \caption{$K^{*0}$ and $\phi$ to $K^{-}$ ratios as function of multiplicity from RHIC (left)~\cite{SQM3} and LHC (right)~\cite{kphi}.}
    \label{Kphi}
  \end{center}
\end{figure}

Lifetimes of long-lived resonances are comparable to the typical lifetime of the QGP fireball created in heavy-ion collisions.
Resonances can thus be used to study the properties and evolution of the hot and dense QGP medium.
The $K^{*0}$ and $\phi$ mesons have different hadronic cross sections and lifetimes. 
The comparison of $\phi/K^-$ and $K^{*0}/K^-$ ratios in Fig.~\ref{Kphi} indicate strong medium effects at RHIC and LHC~\cite{SQM3,kphi}. 

Dileptons are penetrating probe to heavy-ion collisions~\cite{diel}.
Recent measurements show a strong enhancement in the very low $\pT$ region. 
The results point to additional physics contributions,
for example contributions from photon interactions in the 
magnetic field trapped in the QGP~\cite{lowpt}.

\section{Chirality, vorticity and polarization effects}
\begin{figure}[htb]
  \begin{center}
    \includegraphics[width=0.36\textwidth]{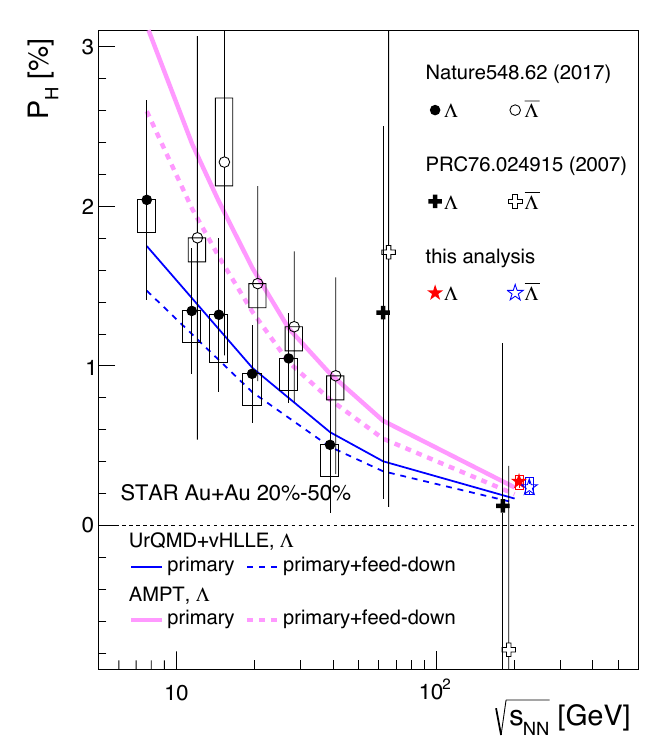}
    \includegraphics[width=0.44\textwidth]{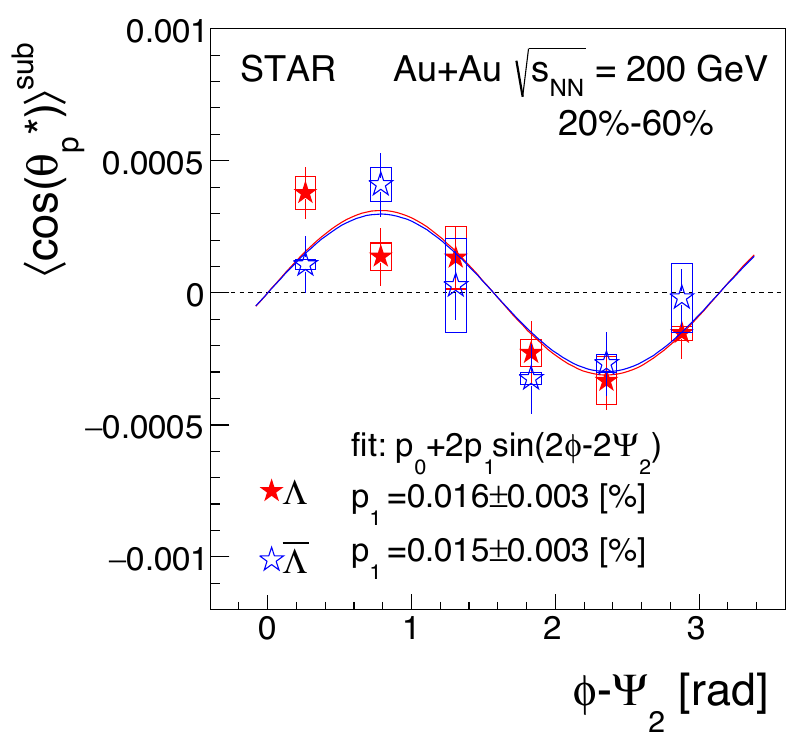}
	  \caption{(Left) Energy dependence of the global polarization of $\Lambda$ and $\bar\Lambda$ in Au+Au collisions~\cite{LambdaN}. (Right) local polarization of $\Lambda$ and $\bar\Lambda$ as a function of azimuthal angle~\cite{LambdaP}.} 
    \label{LambdaP}
  \end{center}
\end{figure}

Due to spin-orbit coupling, particles produced in non-central heavy-ion collisions possess large orbital angular momentum and can be globally polarized along the angular momentum direction~\cite{LambdaPRL}.
This effect was demonstrated by the global $\rm{\Lambda}$ polarization measurement from STAR (left panel of Fig.~\ref{LambdaP})~\cite{LambdaN}.
The data also hint a systematic splitting between $\rm{\Lambda}$ and $\rm{\bar{\Lambda}}$, 
an effect expected from the initial magnetic field.
Recently, STAR reported a first observation of the $\rm{\Lambda}$ local polarization with a quadrupole structure (right panel of Fig.~\ref{LambdaP}), which could be related to the elliptic flow~\cite{LambdaP}.

\begin{figure}[htb]
  \begin{center}
    \includegraphics[width=0.44\textwidth]{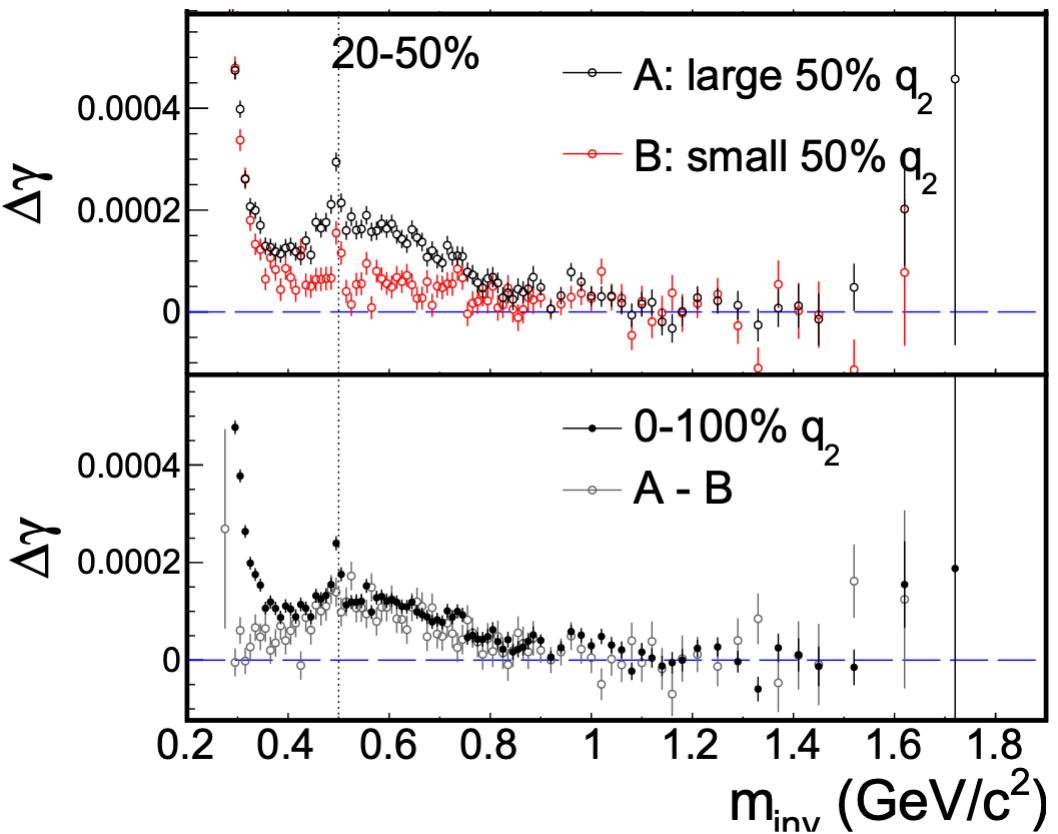}
    \includegraphics[width=0.40\textwidth]{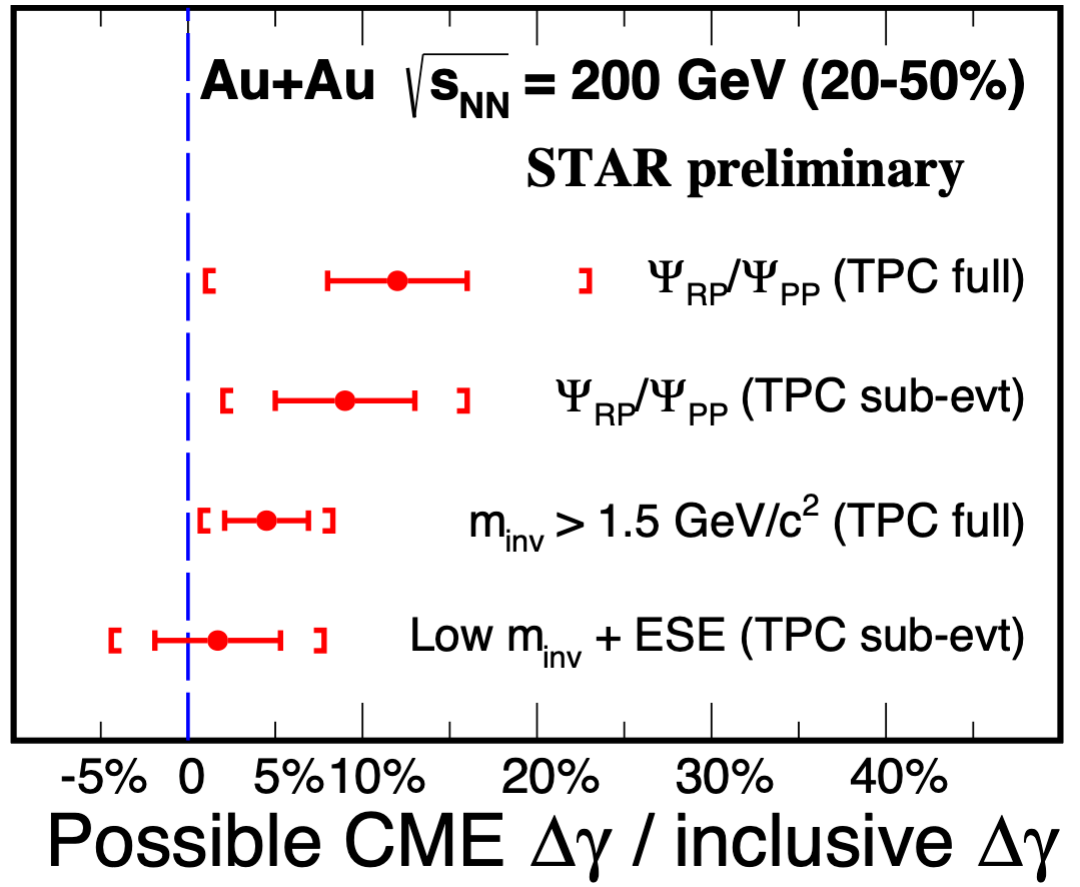}
	  \caption{(Left) $\Delta\gamma$ correlator as function of invariant~\cite{CMEjie}. (Right) relative contribution from possible CME signal to the measured $\Delta\gamma$~\cite{CMEjie,PPRP,massEPJC}.} 
    \label{CME}
  \end{center}
\end{figure}

An electric charge separation can be induced by chirality imbalance along the strong magnetic field and is predicted to occur in relativistic heavy-ion collisions because of topological charge fluctuations and the approximate chiral symmetry restoration in QCD. 
This effect is called the Chiral Magnetic Effect (CME)~\cite{PPNP}. 
Since the first measurement of the $\Delta\gamma$ correlator in 2009~\cite{CME09},
there have been extensive developments to reduce or eliminate the backgrounds~\cite{PPNP}.
Figure~\ref{CME} (left) shows the results obtained using the invariant mass method, one of the recently developed method~\cite{CMEjie}. 
The extracted potential CME signal relative to the inclusive $\Delta\gamma$ 
in 200 GeV Au+Au collisions with two novel methods~\cite{PPRP,massEPJC} are summarized in the right panel of Fig.~\ref{CME}.
These data-driven estimates indicate that the possible CME signal is small, within 1-2 $\sigma$ from zero~\cite{CMEjie}.


\section{Summary}
The recent results on light flavor from the STAR experiment are overviewed. 
The longitudinal flow decorrelation was measured in heavy-ion data and compared to LHC data. 
The elliptic anisotropy is measured \pAu\ and \dAu\ collisions. 
These measurements will further our understanding of the importance of the initial geometry to the system evolution. 
The net-proton (net-$\Lambda$) cumulants,
the light nuclei coalescence parameter and neutron density fluctuation
are reported. All these results seem to show non-monotonic behavior with collision energy and may bear important implications to phase transitions and the possible critical point.
The strange hadron production is found to be not suppressed at \sNN $\leq$ 11.5 GeV, calling for further studies at low energies.
The measured ratios of resonance yields to $K^{-}$ indicate strong medium effects. 
Strong enhancement is observed in the very low $\pT$ dielectron yield, which may be due to photon interactions.
Hypertriton measurements are reported, which present the first test of the CPT symmetry in the light hypernuclei sector.  
The $\Lambda$ local polarization with a quadrupole structure is observed for the first time,
which needs further theoretical undertanding.
Two novel data-driven methods are used to search for the CME signal. 
The present estimates indicate that the possible CME signal is small, within 1-2 $\sigma$ from zero.


\paragraph{Acknowledgements}
This work was supported by the U.S. Department of Energy (Grant No. de-sc0012910).



%
%
%
%
%
%
%
%
\end{document}